\documentclass[twocolumn,preprintnumbers,amsmath,amssymb,apl,superscriptaddress]{revtex4-1}

\usepackage{graphicx}
\usepackage{dcolumn}
\usepackage{bm}

\def\opex{ Opt.\ Express }
\def\ao{ Appl.\  Opt.\ }

\def\apl{ Appl.\ Phys.\ Lett.\ }

\def\josab{ J.\ Opt.\ Soc.\ Am.\ B }

\def\ol{ Opt.\ Lett.\ }

\def\prl{ Phys.\ Rev.\ Lett.\ }

\begin{document}


\title{Dispersion-free monochromatization method \\for selecting a single-order harmonic beam}

\author{Eiji J. Takahashi}
 \email{ejtak@riken.jp}
\affiliation{%
Extreme Photonics Research Group, RIKEN Center for Advanced Photonics, 2-1 Hirosawa, Wako, Saitama 351-0198, Japan
}%

\author{Masatoshi Hatayama}%
\affiliation{%
NTT Advanced Technology Corporation, 3-1 Morinosato Wakamiya, Atsugi, Kanagawa, 243-0124, Japan 
}%

\author{Satoshi Ichimaru}%
\affiliation{%
NTT Advanced Technology Corporation, 3-1 Morinosato Wakamiya, Atsugi, Kanagawa, 243-0124, Japan 
}%

\author{Katsumi Midorikawa}%
\affiliation{%
Extreme Photonics Research Group, RIKEN Center for Advanced Photonics, 2-1 Hirosawa, Wako, Saitama 351-0198, Japan
}%

%

\date{\today}

\begin{abstract}
We propose a method to monochromatize multiple orders of high harmonics by using a proper designed multilayer mirror.
Multilayer mirrors designed by our concept realize the perfect extraction of a single-order harmonic from multiple-order harmonic beam,
and exhibit broadband tunability and high reflectivity in the soft-x-ray region.
Furthermore, the proposed monochromatization method can preserve the femtosecond to attosecond pulse duration for the reflected beam. 
This device is very useful for ultrafast soft x-ray experiments that require high-order harmonic beams, such as femtosecond/attosecond, time-resolved, pump-probe spectroscopy.
\end{abstract}

\pacs{Valid PACS appear here}
\maketitle

High-order harmonic generation (HHG) can produce  coherent soft-x-ray bursts that last only a few hundred attoseconds. 
To date, the pulse duration of an isolated attosecond pulse reaches sub-100 as region \cite{80as}, and its pulse energy reaches as high as microjoule level \cite{tak_NC}. 
By the way, tunable coherent soft-x-ray  pulses are required for applications such as an ultrafast pump-probe spectroscopy (see review ref. \cite{km}). 
Although HHG can produce femtosecond/attosecond, coherent beams in the soft-x-ray region, time-resolved pump-probe experiments are best performed with tunable monochromatized radiation. 
Thus, monochromatization with tunability is a key subject for time-resolved pump-probe experiments that use high-order harmonic (HH) beams.
However, monochromatization should be achieved in a manner that preserves an ultrashort pulse duration of an HH beam. 

One proposition to extract a single-order  HH beam and preserve its pulse duration is the time-delay compensated monochromator (TDCM) \cite{TDCM1,TDCM2,TDCM3,TDCM4,TDCM6,TDCM7}, which comprises a pair of gratings that compensate for the pulse front tilt. 
However, one drawback of the TDCM is a low throughput, which results from the low diffraction efficiency of the gratings. 
Moreover, TDCM requires not only a complex optical configuration but also temporal characterization \cite{TDCM3} of the extracted HH beam to confirm the compensation of the pulse front tilt.
Furthermore, it is extremely  difficult to completely compensate for the stretched pulse by employing the second grating \cite{TDCM6,TDCM7}.
Another monochromatization method uses a time-compensated monochromator with a multilayered mirror (MLM) \cite{Poletto}.
However, 
a traditionally  designed MLM exhibits a long reflective tail that depends on the incident angle, and has a second-order reflectivity in the higher photon energy region.   
Thus, it is quite difficult to extract a single-order HH beam in a wide tuning range.

%

In this paper, we propose a properly  designed MLM monochromator to efficiently extract a single-order HH beam.
By considering a fabrication procedure for MLMs, we have established a design rule for monochromatizing multiple orders of HH beams using MLMs.
Not only is the proposed monochromatization method easy to perfoerm, but it also perfectly isolates a single-order HH  beam in the 25 - 130 eV region. 
Moreover, this monochromatization method preserves the femtosecond to attosecond temporal duration in the extracted HH beam because of the low dispersion value of our designed MLM.
This device can provide wide photon energy tunability on the femtosecond/attosecond, time-resolved, pump-probe experiment using HH beams.
In addition, our method is useful for applications using HH beam, such as a seeded FEL \cite{togashi,seedFEL}, and imaging experiments \cite{ime1,ime2}.






First, we examine the designing step of an MLM for extracting a single-order HH beam. 
Fabrication of a high-reflectivity EUV MLM requires (i) a large reflective index difference between low-$Z$ material and high-$Z$ material, (ii) a small periodic length  fluctuation, and (iii) a smooth interface. 
Periodic length of multilayer and peak wavelength are related to Bragg's Law: $nd=m\lambda / (2\sin\theta)$.
Here the peak wavelength is shown as a function of $m$, $\theta$, and $d$: $\lambda (m, \theta, d) = 2 d \sin \theta/m$,
where $n$ is an actual reflective index, $d$ is a periodic length of multilayer, $\theta$ is a grazing incident angle, and $m$ is an order of Bragg's reflection. 
The real reflectivity ($R$) of an MLM is related to interface roughness as a Debye-Waller Factor; $\exp \left(-2 (2 \pi \sigma \cos \theta/\lambda)^2 \right)$, where $\sigma$ is the root mean square interface roughness.
It is also known that a narrow-bandwidth EUV mirror is fabricated using a small $\gamma$ (ratio of high-$Z$ material thickness to $d$) multilayer \cite{ml1}.
Thus, to design the multilayer for monochromatization of an HH beam, it is most important to select a suitable low-$Z$ material, as the optical constants of a low-$Z$ material largely contributes to optical characteristics, particularly variation in reflective profiles at each incident angle.
As a first step in the multilayer design of our monochromator, a low-$Z$ material must be selected, which has a considerably smaller optical constant variation and absorption coefficient at the target energy range. 
In the second step, periodic length $d$ is determined such that $\lambda(2, 90, d) = nd$ is smaller than the absorption edge of the low-$Z$ material. 
Thus, reflectivity from the first order diffraction of the multilayer becomes considerably much higher than that from the higher-order diffraction. 
From these steps, the energy range for the MLM is determined.
In the third step, a suitable high-$Z$ material is selected based on both its reflectivity and multilayer fabrication properties, as periodic fluctuations and interface roughness strongly influence the quality of an MLM.
As a final step, $\gamma$ and a layer number ($N$) are optimized to suit the bandwidths and photon energy intervals for the specific HHG spectrum. 

\begin{table}
  \centering
  \caption{Parameters for multilayer mirrors used as a monochromator. 
  }\begin{tabular}{cccc} \\ \hline \hline
     & ~~Period: $d$ (nm) & ~~~Bottom (nm) & ~~~Layers: $N$ \\ \hline
    SiC/Mg & 24.00   & Mg: 21.60 & 60  \\
    Zr/Al & 16.00   & Al: 14.40 & 60 \\
    Mo/Y & 9.00   & Y: 8.10  & 80  \\ \hline \hline
  \end{tabular}
\end{table}

\begin{figure}
\centerline{\includegraphics[width=9.2cm]{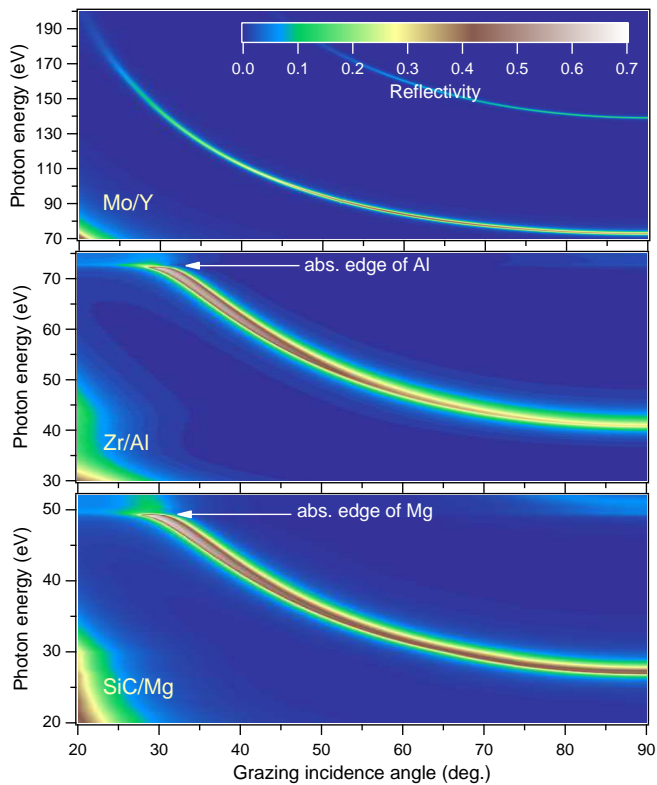}}
\caption{Reflectivity of a multilayer mirror. Top: Mo/Y, Middle: Zr/Al, Bottom: SiC/Mg.}
\label{ref}
\end{figure}

Commonly, almost all the HHG experiments are demonstrated using a multicycle Ti: sapphire laser at 800 nm ($\sim$ 1.55 eV), and consequently, HHG produces a discrete spectrum with photon energy intervals of  $\sim$ 3.1 eV. 
As a general case, we designed MLMs for extracting a single-order HH driven by a 1.55 eV pump. 
Table 1 shows the design parameters of three MLMs for our monochromator.
To effectively isolate a single-order HH with specific photon energy in a reflected beam, one must optimize the multilayer parameters, as mentioned above. 
Here we explain the practical SiC/Mg multilayer designing procedure in detail. 
Note that Zr/Al and Mo/Y are also designed according to the same policy. 
In the low photon energy region, Mg, Al, and Si become candidates as low-Z materials. 
Due to its low absorption coefficient, small variation in refractive index \cite{palik}, and attenuation of a higher-order reflection, we chose Mg as the low-Z material. 
Next, we examined a suitable high-$Z$ material for layering with Mg. 
Due to their reflectivity properties, SiC, B$_4$C, Mo, and Pt are potential candidates as high-$Z$ materials. 
Since Pt and Mo, like Mg, are metals, the interface roughness of a multilayer will be increased because of instability of the layer. 
Although the B$_4$C has a slightly higher reflectivity as compared to SiC, we selected SiC as the high-$Z$ material because of previous manufacturing results \cite{ml2,ml3}.
Finally, $N$ and $\gamma$ were decided to be 60 and 0.1, respectively, to isolate a single-order HH beam driven by the 1.55 eV pump. 
Similarly, Zr/Al and Mo/Y (see Table 1)  were designed for photon energies of 40 - 70 eV and 70 - 130 eV, respectively. 
Note that these material combinations are also motivated by considering multilayer fabrication.

\begin{figure}
\centerline{\includegraphics[width=9cm]{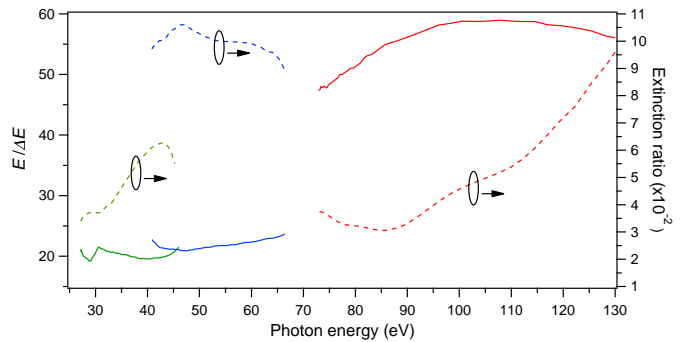}}
\caption{Energy resolution and extinction ratio as a function of photon energy. Green: SiC/Mg, Blue: Zr/Al, Red: Mo/Y.
}
\label{ER}
\end{figure}

\begin{figure}[htb]
\centerline{\includegraphics[width=9cm]{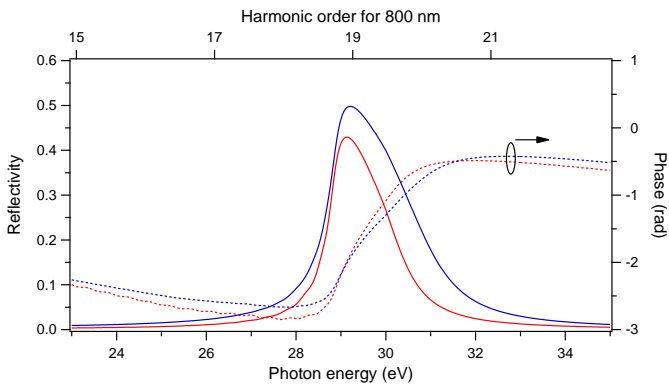}}
\caption{Calculated reflectivity of SiC/Mg multilayer mirror at $\theta_{\rm SiC/Mg} = 70^\circ$ and its spectral phase. Red profile: design parameters shown in Table. 1. Blue profile: Mg thickness is 20.40 nm, but other parameters are the same as those in Table. 1
}
\label{atto_chirp}
\end{figure}

The bottom panel in Fig. \ref{ref} shows the reflectivity of SiC/Mg MLM as a function of incident angle $\theta$.
Here, $\sigma$ was fixed to 0.
The reflectivity curve for the $s$-polarized light shows a single peak with a Gaussian-like profile (see also Fig. \ref{atto_chirp}), and its profile is almost  maintained within the tuning range.
Although a small portion of the reflectivity, due to $m=2$, appears at higher than the $\sim$ 50 eV region,  the peak reflectivity at 27 eV is almost 10 times higher than that of $m=2$ due to the absorption effect of Mg.
As is shown, the reflected photon energy gradually changes, depending on the incident angle, while keeping a single peak profile.
This corresponds to the spectral selection of the single HH beam by changing the incident angle ($\theta$).
Moreover, the average reflectivity stays 50 \% in the tunable range.
The middle and top panels in Fig. \ref{ref} show the reflectivity of Zr/Al and Mo/Y.
Zr/Al and Mo/Y can operate within the photon energy range of 40 eV - 70 eV with an average reflectivity of 40 \%, and within the range of 70 eV - 130 eV with an average reflectivity of 30 \%,  respectively.
Unfortunately, Mo/Y has a relatively high second-order reflectivity because Y does not exhibit an absorption edge at 130 eV. 
When HHG of the 1.55-eV pump is demonstrated using rare gas targets, except He, the second-order reflectivity of Mo/Y does not cause a major problem because the cutoff energy is lower than 130 eV \cite{80as,tak6}.
Note that a broadband reflectivity gradually appears at an incident angle below $\sim 30^\circ$ because a total reflection occurs at the top layer. 
Thus, the lowest incident angle of our MLM is dependent on a total reflection.



Figure \ref{ER} shows the energy resolution of an MLM, which is defined by $E/\Delta E$, where $\Delta E$ is a bandwidth (FWHM) of the reflective photon energy at each incident angle. 
In the tuning range, $\Delta E$ of SiC/Mg, Zr/Al and Mo/Y change from 1.3 eV to 2.2 eV, 1.8 eV to 2.8 eV, and 1.5 eV to 2.3 eV, respectively.
Thus, $E/\Delta E$ shows a relatively flat profile  in the tunable range.
As is discussed below, the energy resolution relates to the acceptable pulse duration for the reflected beam.
Next, to evaluate the efficiency of the single HH beam isolation, we calculated an extinction ratio, defined by  $(R( E + 3.1 {\rm eV}) + R( E -3.1 {\rm eV}) ) /2 R_{\rm max}(E)$;  $R( E \pm 3.1 {\rm eV})$, which corresponds to the reflectivity of both sides of harmonic order.
This index is the averaged value at both sides because a reflective profile is not perfectly symmetric as a function of the photon energy (see Fig. \ref{atto_chirp}). 
As is shown in Fig. \ref{ER}, an extinction ratio almost maintains $ \sim 10^{-2}$ order  in the tunable range.
This value is sufficient for isolating a single-order HH beam; however, we can improve the extinction ratio by two orders of magnitude by employing a pair of MLMs.

To discuss the acceptable pulse duration for the reflected beam, we evaluated the dispersion of the multilayer coating. 
Though our designed MLM has enough bandwidth for reflecting sub-femtosecond pulses, a small amount of dispersion strongly influences the pulse duration if its temporal duration is on the order of attoseconds. 
The calculated reflectivity of the SiC/Mg MLM at $\theta$ = 70$^\circ$ is shown by the solid red line in Fig. \ref{atto_chirp}, and its calculated spectral phase is given by the dashed line in the same figure. 
As is shown, the reflectivity of SiC/Mg has a Gaussian-like profile with an exponential tail.
Since the propagation length of light inside the layer depends on the incident angle, the optical dispersion of our MLMs also changes, depending on the incident angle.
The optical dispersion gradually decreases with the incident angle.
$\Delta E$ becomes  $\sim$ 1.5 eV at $\theta$ = 70$^\circ$,  which can support a $\tau_0 = $1.22 fs transform-limited (TL) pulse duration, 
according to $\tau_0 ({\rm as}) = 1835/ \Delta E$ with a Gaussian profile. 
When we consider the dispersion from the multilayer coating, the duration of the reflected pulses is 1.25 fs. 
Thus, we can safely conclude that our MLM almost preserves the TL pulse duration on the reflected beam.
In addition, we can support attosecond pulse duration by optimizing the multilayer design.
The solid blue and dashed blue profiles correspond to  the reflectivity and the spectral phase of another SiC/Mg multilayer at $\theta$ = 70$^\circ$, which was designed for covering an attosecond duration.
Here, we changed the Mg thickness from 21.60 nm to 20.40 nm, while other parameters fixed.
$\Delta E$ increases to 2.0 eV, which supports a 910 as TL pulse duration. 
Although dispersion occurs, the duration of the reflected pulses is 935 as. 
Both of maximum reflectivity and bandwidth increases; however, the extinction ratio slightly decreases from $2.6 \times 10^{-2}$ to $4.6 \times 10^{-2}$.
Note that the energy resolution and the extinction ratio are in the relationship of trade off.
The designed Zr/Al and Mo/Y layers also exhibit similar temporal characteristics for the reflected beam.
These mirrors  can support a femtosecond, as well as an attosecond pulse duration, depending on the designed energy resolution.



\begin{figure}[htb]
\centerline{\includegraphics[width=9cm]{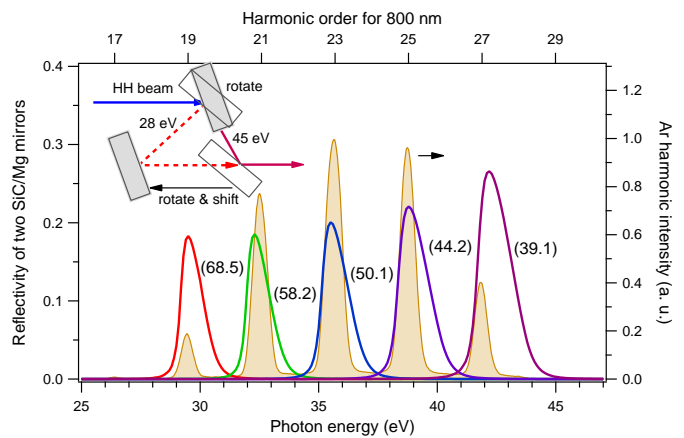}}
\caption{Reflective profiles of a pair of SiC/Mg multilayer mirrors as a function of photon energy. Filled profile shows the HH spectrum  from Ar. Each number in parentheses refers  to the incident angle ($\theta_{\rm SiC/Mg}$). Inset: A typical setup of our monochromator.
}
\label{SiCMg}
\end{figure}

In the inset of Fig. \ref{SiCMg}, we show an example setup of a monochromator using two MLMs, which is similar to that of a monochromator for synchrotron orbit radiation \cite{SOR}.
Although the total throughput is decreased, it is possible to fix the beam propagation direction using the pair of MLMs. 
We show the typical reflective profiles of a pair of SiC/Mg MLMs at several incident angles. 
The amber-filled profile exhibits a high-order harmonic spectrum from Ar \cite{tak_PRL}, which is generated by a 30 fs Ti:sapphire laser. 
A pair of SiC/Mg MLMs attains $\sim$ 20 \% reflectivity,  which is equivalent to that of a six-optics setup of TDCM \cite{TDCM8}.
Here, the $E/\Delta E$ slightly increases to $\sim$ 25, owing to the narrowing of the reflected beam, and the extinction ratio reaches $\sim$ $10^{-4}$.
By simply rotating the MLMs, we can precisely choose and isolate a single-order HH as the reflected beam. 
Figure 5 shows reflective profiles of a pair of Zr/Al and Mo/Y MLMs.
Here each reflective profile corresponds to the incident angle adjustment for isolating a single odd-order harmonic beam (see top axis). 
As is shown, Zr/Al and Mo/Y MLMs well work as monochromators for isolating a single-order HH beam.
$E/\Delta E$ is to be $\sim$ 25 in Zr/Al and $\sim$ 60 in Mo/Y.
Note that one drawback of our monochromatization method is the lack of separation of the pump beam.
For eliminating the pump beam, we will need to install beam splitters \cite{tak_BS,nagata} or a proper metal filter.



\begin{figure}
\centerline{\includegraphics[width=9cm]{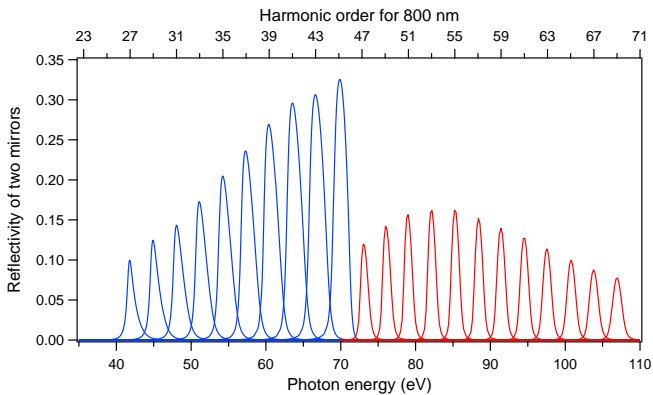}}
\caption{Reflective profiles of a pair of Zr/Al (blue) and Mo/Y (red) mirrors as a function of photon energy. Incident angle: $\theta_{\rm Zr/Ar}$ = 78$^\circ$ - 33$^\circ$, $\theta_{\rm Mo/Y}$ = 86$^\circ$ - 43$^\circ$.
}
\end{figure}


In conclusion, we propose a monochromatization method that uses a proper designed MLM to efficiently extract a single-order HH beam. 
Our method is easy to handle and exhibits broadband tunability in the soft x-ray region. 
By simply rotating the MLMs, we can continuously choose and precisely isolate a single-order HH as the reflected beam. 
In our monochromator, we can obtain wide tunability in the photon energy range of 27 to 130 eV using an MLM with spatially separated coatings of SiC/Mg, Zr/Al, and Mo/Y.
Moreover, this monochromatization method preserves the femtosecond to attosecond temporal duration in the extracted HH beam. 
At present, a super continuum HH beam in the soft x-ray region is demonstrated for generating an isolated pulse with $<$ 100 as \cite{80as}.
By combining a super continuum HH beam and our monochromator,  we can realize a tunable, coherent soft x-ray source with a sub-femtosecond pulse duration.
Thus, our method will be very useful for ultrafast soft x-ray experiments such as femtosecond/attosecond, time-resolved, pump-probe spectroscopy. 


This work was supported by the Ministry of Education, Culture, Sports, Science and Technology (MEXT) through a Grant-in-Aid for Scientific Research (B) No. 25286074, and through a Grant-in-Aid for Exploratory Research No. 24656069.

\end{document}